\documentclass[preprint]{sig-alternate}

\conferenceinfo{SIN'10,} {Sept. 7--11, 2010, Taganrog, Rostov-on-Don, Russian Federation.}  
\CopyrightYear{2010} 
\crdata{978-1-4503-0234-0/10/09} 
\usepackage{multirow}

\toappear{Permission to make digital or hard copies of all or part of this work for personal or classroom use is granted without fee provided that copies are
not made or distributed for profit or commercial advantage and that copies
bear this notice and the full citation on the first page. To copy otherwise, to
republish, to post on servers or to redistribute to lists, requires prior specific permission and/or a fee.\\
This is a preprint version of our paper presented in SIN'10, Sept. 7-11, 2010, Taganrog, Rostov Oblast, Russia.}

\title{Ontological Approach toward Cybersecurity\\in Cloud Computing}

\numberofauthors{3}
\author{
\alignauthor Takeshi Takahashi\\
\affaddr{National Institute of Information and Communications Technology}\\
\affaddr{4-2-1 Nukui-Kitamachi Koganei}\\
\affaddr{Tokyo Japan}\\
\email{takeshi\_takahashi@nict.go.jp}
\alignauthor Youki Kadobayashi\\
\affaddr{Nara Institute of Science and Technology}\\
\affaddr{Takayama 8916-5, Ikoma}\\
\affaddr{Nara Japan}\\
\email{youki-k@is.naist.jp}
\alignauthor Hiroyuki Fujiwara\\
\affaddr{Solution Crew Inc.}\\
\affaddr{Shin-osaka Wako Bldg. 8F 4-6-18 Miyahara Yodokawa-ku}\\
\affaddr{Osaka Japan}\\
\email{reisei@scw.co.jp}
}

\begin{document}

\maketitle

\begin{abstract}

Widespread deployment of the Internet enabled building of an emerging IT delivery model, i.e., cloud computing.
Albeit cloud computing-based services have rapidly developed, their security aspects are still at the initial stage of development.
In order to preserve cybersecurity in cloud computing, cybersecurity information that will be exchanged within it needs to be identified and discussed.
For this purpose, we propose an ontological approach to cybersecurity in cloud computing.
We build an ontology for cybersecurity operational information based on actual cybersecurity operations mainly focused on non-cloud computing.
In order to discuss necessary cybersecurity information in cloud computing, we apply the ontology to cloud computing.
Through the discussion, we identify essential changes in cloud computing such as data-asset decoupling and clarify the cybersecurity information required by the changes such as data provenance and resource dependency information.

\end{abstract}





\newpage


\section{Introduction}\label{Sec:introduction}

Information technology is rapidly evolving.
Widespread deployment of the Internet enabled construction of an emerging IT delivery model, i.e., cloud computing.
Albeit there exist various definitions of cloud computing, one of the commonly recognized definition is provided by the National Institute of Standards and Technology (NIST): Cloud computing is a model for enabling convenient, on-demand network access to a shared pool of configurable computing resources (e.g., networks, servers, storage, applications, and services) that can be rapidly provisioned and released with minimal management effort or service provider interaction \cite{NIST_CloudDefinition}.
Cloud computing is massively scalable, provides a superior user experience, and is characterized by new, Internet-driven economics. 

Following the emergence of cloud computing, services over cloud computing, i.e., cloud services, have been rapidly developed.
Cloud services, such as Amazon Web Services and Google Apps, are accessible via a web browser or web service application programming interface (API) and are regarded as convenient and cost-saving.
The market size of cloud services, in terms of spending, is therefore growing rapidly; with the USD 17 billion in 2009 expected to burgeon into USD 44 billion in 2013 \cite{IDC_CloudForecast}.
In terms of the percentage of the total IT market size, IT cloud services are expected to grow from 5\% in 2009 to 10\% in 2013 \cite{IDC_CloudForecast}, which means they will outpace traditional IT spending over the coming years.

Cloud services are, however, provided by individual cloud service operators and have very little interoperability.
In order to build and secure the interoperability, it is necessary to build international standards, which improve application portability enabling resource accommodation between cloud service providers.
With these advancements, reliability in times of disaster can be drastically improved.
Major organizations such as Open Grid Forum (OGF), Distributed Management Task Force (DMTF) and Storage Network Industry Association (SNIA) are currently focusing on the service interoperability issues \cite{DMTF_WhitePaper,CDMI,OCCi_requirements}.
Regarding security issues, albeit their importance is advocated and some security guidances are provided by the Cloud Security Alliance (CSA) \cite{CSA_guidance}, technical standard-building in cloud computing is still in its initial stages of development.

Preserving cybersecurity in cloud computing requires identifying what kind of cybersecurity information needs to be exchanged.
For this purpose, we propose an ontological approach.
We build an ontology for cybersecurity operational information based on actual cybersecurity operations mainly focused on non-cloud computing.
In order to discuss necessary cybersecurity information in cloud computing, we apply the ontology to cloud computing.
The ontology shows a holistic perspective of cybersecurity operations and provides the categories of cybersecurity operational information.
For each of these, we discuss cybersecurity information that is newly required or that needs modifying to suit a cloud computing context.
For instance, we discuss what type of incident log will be required for incident handling operations in cloud computing, and what type of asset description information will be required for managing IT assets for each user organization in cloud computing.
Through this discussion, we identify essential changes in cloud computing such as data-asset decoupling and clarify the cybersecurity information necessitated by changes such as data provenance and resource dependency information.

The rest of this paper is organized as follows: 
section \ref{Sec:RelatedWorks} presents related works,
section \ref{Sec:Ontology} describes our proposed ontology,
section \ref{Sec:Discussion} discusses cybersecurity in cloud computing,
section \ref{Sec:EssentialChanges} summarizes essential changes in cloud computing extracted from the discussion and section \ref{Sec:Conclusion} concludes this paper.

\section{Related Works}\label{Sec:RelatedWorks}

In order to build an ontology for cybersecurity operational information, it is beneficial to understand the need for an ontology and some security ontology works.

An ontology is an explicit specification of a conceptualization, which is an abstract, simplified view of the world that we wish to represent for certain purposes \cite{OntologyPrinciples}.
Ontologies are useful as means to support sharing and reutilization of knowledge \cite{OntologyDecker}.
This reusability approach is based on the assumption that if a modeling scheme, i.e., an ontology, is explicitly specified and mutually agreed upon by the parties involved, then it is possible to share, reutilize and extend knowledge \cite{Tsoumas_Springer2005}.

Fenz et al. proposed a security ontology \cite{Fenz_AsiaCCS2009} with concepts grouped into three subontologies: security, enterprise, and location.
The security subontology introduces five concepts: attribute, threat, rating, control and vulnerability.
Wang et al, introduced an ontology for security vulnerabilities \cite{OVM,OVM_ieee}, which focuses on software vulnerabilities and discussed the National Vulnerability Database (NVD) \cite{NVD}.
Tsoumas et al. extended the DMTF Common Information Model (CIM) standard with ontological semantics in order to utilize it as a container for IS security-related information, and proposed an ontology of security operation for an arbitrary information system and defined it in OWL \cite{Tsoumas_Springer2005,Tsoumas_AINA2006}.
Parkin et al. proposed an information security ontology incorporating human-behavioral implications \cite{Simon_SIN2009}.
This ontology provides a framework for investigating the casual relationships of human-behavioral implications resulting from information security management decisions, before security controls are deployed.
Denker et al. proposed several ontologies for security annotations of agents and web services using OWL \cite{Denker_Report}.
They mainly addressed knowledge representation and some of the reasoning issues for trust and security in the Semantic Web.
Albeit there exist various other ontology works, the reusability of their ontologies is rather limited or they are still at early stages of development, as Blanco et al. discussed in a survey of security ontologies \cite{Blanco_2008}.


Different from the aforementioned works, our viewpoint is on actual cybersecurity operations, and we focus on building an ontology of cybersecurity operational information.
For practicality and reusability, we build the ontology based on intensive discussions with cybersecurity operators.
The ontology can provide a framework for sharing and reutilizing cybersecurity operational information and can define the terminology.
Our initial work is found in \cite{Take_CSIIRW2010}.


\section{Ontology of Cybersecurity Operational Information}\label{Sec:Ontology}

Based on intensive discussions with major cybersecurity operators, we build an ontology of cybersecurity operational information.
The discussions covered actual cybersecurity operations in the USA, Japan, and Korea.
Albeit each cybersecurity operator runs slightly different operations, we succeeded in building a generalized ontology of cybersecurity operational information.
First, we define the domains for cybersecurity operations in section \ref{Sec:OperationDomains}, and identify the entities required to run the operations in the domains in section \ref{Sec:Entities}. Based on the domains and entities, we identify cybersecurity information provided by entities in each operation domain and build the ontology of cybersecurity operational information in section \ref{Sec:CybersecurityInformation}.


\subsection{Cybersecurity Operation Domains}\label{Sec:OperationDomains}

The term "cybersecurity operation" covers a range of security operations in cyber society.
Nevertheless, this paper focuses on the cybersecurity operations that preserve information security in cyber societies.
Information security is the preservation of information confidentiality, integrity, and availability \cite{OECD_SecurityGuideline}.
Sometimes, it also encompasses accountability, authenticity, and reliability of the information \cite{13335-1}.

Cybersecurity operations consist of three domains: IT Asset Management, Incident Handling and Knowledge Accumulation.

The IT Asset Management domain runs cybersecurity operations inside each user organization such as installing, configuring and managing IT assets.
The IT asset includes both resources of users and providers; it includes not only an user's own IT assets but also network connectivity, cloud services and identity services provided by external entities for the user.

The Incident Handling domain detects and responds to incidents occurring in cyber societies by monitoring computer events, incidents comprised of several computer events, and attack behaviors caused by the incidents.
More specifically, it monitors computer events, and when an abnormality is detected, then it produces an incident report. Based on the report, it investigates the incident in detail so that it can clarify the attack pattern and its countermeasures.
Based on the analysis of incidents, it may provide alerts and advisories, e.g. early warnings against potential threats, to user organizations.


The Knowledge Accumulation domain researches cybersecurity and generates reusable knowledge for other organizations.
For reusability by those organizations, it provides common naming and taxonomy, through which it organizes and accumulates the knowledge.


\subsection{Entities}\label{Sec:Entities}

Based on the cybersecurity operation domains defined in section \ref{Sec:OperationDomains}, this section identifies entities necessary for running cybersecurity operations in each domain.
Note that the entities are defined from the viewpoint of functions; therefore one instance of an entity may be an instance of another entity in the real world.

In the IT Asset Management domain, there exist two entities for its operation: Administrator and IT Infrastructure Provider.
The Administrator administers the system of its organization, possessing information on its own IT assets.
The system administrator inside each organization is its typical instance.
The IT infrastructure Provider provides each organization with IT infrastructure, which includes the network connectivity, cloud services such as software as a service (SaaS), platform as a service (PaaS) and infrastructure as a service (IaaS), and identity.
The Internet Service Provider (ISP) and Application Service Provider (ASP) are its typical instances.

In the Incident Handling domain, there exist two entities for its operation: Response Team and Coordinator.
Response Team is an entity that monitors and analyzes various kinds of incidents in cyber-societies, e.g. unauthorized access, DDoS attacks and phishing, and accumulates incident information.
Based on the information, it may implement countermeasures, e.g. registering phishing site addresses on black lists. The incident response team inside a Managed Security Service Provider (MSSP) is its typical instance.
The Coordinator is an entity that coordinates with the other entities and addresses potential threats based on known incidents and crime information.
The CERT Coordination Center (CERT/CC), be it either commercial or non-commercial, is its typical instance.

In the Knowledge Accumulation domain, there exist three entities for its operation: Researcher, Product \& Service Provider and Registrar.
The Researcher is an entity that researches cybersecurity, extracts knowledge from the research and accumulates it.
Cybersecurity research teams inside MSSP, e.g. X-force within International Business Machines Corp. (IBM) as well as the Risk Research Institute of Cyber Space at the Little eArth Corporation Co., Ltd. (LAC), are its typical instances.
The Product \& Service Provider is an entity that possesses information on products and services, e.g. naming, versions, their vulnerabilities, their patches and configuration information.
A software house, ASP and individual private software programmer are its typical instances.
The Registrar is an entity that classifies, organizes and accumulates cybersecurity knowledge provided by the Researcher and the Product \& Service Provider so that the knowledge can be reutilized by another organization. NIST and the Information-Technology Promotion Agency, Japan (IPA) are its typical instances.


\subsection{Cybersecurity Information}\label{Sec:CybersecurityInformation}

\begin{figure*}[htb]
\begin{center}
\includegraphics[scale=.415]{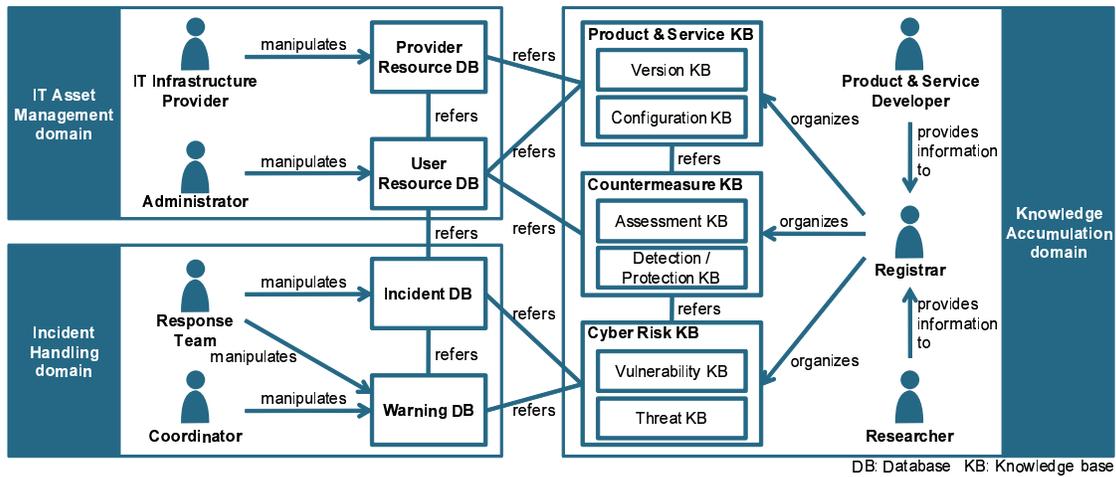}
\caption{Proposed ontology of cybersecurity operational information}
\label{ProposedOntology.eps}
\end{center}
\end{figure*}

Based on the aforementioned operation domains and entities, this section identifies cybersecurity operational information.
Considering the information each of the entities provides, we define four databases and three knowledge bases, as described in the following subsections.

\subsubsection{Incident Database}\label{Sec:IncidentDatabase}

The Incident Database is a database that contains incident-related information.
The Response Team manipulates such information.
Major items of the information stored in this database are the event record, incident record and attack record.

The event record is a record of computer events that includes information on packets, files and their transactions.
Usually, most of the records are provided by computers automatically as computer logs; for instance logs such as log-in time and date as well as terminal information provided when root users log in to a system.
This log is one type of event record.
Common Event Expression (CEE) \cite{CEE} can be utilized to describe the record.

The incident record is a record of incidents, providing description of incidents such as computer states and their consequences.
This record is derived from analyses of several event records and their conjectures, which are created automatically and/or manually.
For instance, when excessive access to one computer is detected, the state of the computer, i.e., excessive access, as well as its expected consequence, i.e., denial of service, should be recorded in an incident record.
Based on this record, the harmfulness of the incident as well as the need for countermeasures can be judged.
Note that an incident record may record false incidents, i.e., incident candidates judged as non-incidents after an investigation.
The Incident Object Description and Exchange Format (IODEF) \cite{IODEF} can be utilized to describe the record.

The attack record is a record of attacks derived from analyses of incident records.
It describes the attack sequence; such as how the attack was initiated, which IT assets were targeted, and how the attack's damage was propagated.

\subsubsection{Warning Database}

The Warning Database is a database that contains information on cybersecurity warnings.
The Response Team and the Coordinator manipulate such information.
These warnings are based on the Incident Database and Cyber Risk Knowledge Base.
Based on the warnings, user organizations may implement countermeasures for warned cybersecurity risks.

\subsubsection{User Resource Database}\label{Sec:UserResourceDatabase}

The User Resource Database is a database that accumulates information on assets inside individual organizations and contains information such as the list of software/hardware, their configurations, status of resource usage, security policies including access control policies, security level assessment result, and intranet topology.
The Administrator manipulates such information.

The Assessment Results Format (ARF) \cite{ARF} and Common Result Format (CRF) \cite{CRF} can be utilized to describe the IT asset assessment results while the Common Vulnerability Scoring System (CVSS) \cite{CVSS}/Common Weakness Scoring System (CWSS) \cite{CWSS} can be utilized to score the IT asset's security level.

This database, as discussed later in section \ref{Sec:CloudServiceSubscriptionInformation}, also contains cloud service subscription information that the individual user organization is utilizing, such as the list of subscribing cloud services (e.g., data center and SaaS) and the usage record of the services.

\subsubsection{Provider Resource Database}

The Provider Resource Database is a database that accumulates information on assets outside individual organizations.
IT Infrastructure Provider manipulates such information.
Two main components of the database are external network information and external cloud service information.

External network information is on networks with which each organization is connected with other organizations such as inter-organization network topology, routing information, access control policy, traffic status and the security level.

External cloud service information includes the service specifications, workload information and security policy information of each cloud service.
Note that, user organization specific information such as local configuration information of each cloud service is stored in the User Resource Database.

\begin{table*}[tb]
\centering
\caption{Major cybersecurity information standards}\label{Tabl:MajorCybersecurityInformationStandards}
\begin{tabular}{p{75pt}|p{65pt}l|p{160pt}}

\hline
Operation Domain &\multicolumn{2}{l|}{Information categories} & Major standards\\
\hline
\multirow{2}{75pt}{IT Asset Management}& \multicolumn{2}{|l|}{User Resource Database}&ARF, CRF, CVSS/CWSS scores\\ \cline{2-4}
& \multicolumn{2}{|l|}{Provider Resource Database} &-\\
\hline
\multirow{2}{75pt}{Incident Handling} & \multicolumn{2}{|l|}{Incident Database} &CEE, IODEF \\ \cline{2-4}
&\multicolumn{2}{l|}{Warning Database}	&-\\
\hline
\multirow{6}{75pt}{Knowledge Accumulation} &\multirow{2}{70pt}{Cyber Risk Knowledge Base}	& Vulnerability Knowledge Base&CVE, CWE\\ \cline{3-4}
&& Threat Knowledge Base	& CAPEC, MAEC\\ \cline{3-4}
\cline{2-4}
&\multirow{2}{70pt}{Countermeasure Knowledge Base}	& Assessment Knowledge Base& CVSS/CWSS formula, OVAL, XCCDF\\ \cline{3-4}
					&& Detection/Protection Knowledge Base& -\\
\cline{2-4}
&\multirow{2}{70pt}{Product Knowledge Base}	& Version Knowledge Base& CPE\\ \cline{3-4}
				&& Configuration Knowledge Base& CCE\\
\hline
\end{tabular}
\end{table*}

\subsubsection{Cyber Risk Knowledge Base}\label{Sec:CyberRiskKnowledgeBase}

The Cyber Risk Knowledge Base is a knowledge base that accumulates cybersecurity risk information and includes two knowledge bases: the Vulnerability Knowledge Base and Threat Knowledge Base.

The Vulnerability Knowledge Base accumulates known vulnerability information, which includes naming, taxonomy and enumeration of known software and system vulnerability information as well as vulnerabilities caused by their mis-configuration. It also includes information on human vulnerabilities -- vulnerabilities exposed by human IT users.
In order to describe the contents of the knowledge base, Common Vulnerabilities and Exposures (CVE) \cite{CVE} and Common Weakness Enumeration (CWE) \cite{CWE} can be utilized.

The Threat Knowledge Base accumulates known cybersecurity threat information including attack knowledge and mis-use knowledge.
Attack knowledge is knowledge of attacks including the information on attack patterns, attack tools (e.g. malware) and their trends.
Trend information includes past attack trends in terms of geography and attack target, for instance, and statistical information on past attacks.
Common Attack Pattern Enumeration and Classification (CAPEC) \cite{CAPEC} and Malware Attribute Enumeration and Characterization (MAEC) \cite{MAEC} can be utilized to describe the contents of the knowledge base.
Mis-use knowledge is on mis-uses caused by users' inappropriate usage, which includes both benign and malicious usage.
Benign usage includes mis-typing, mis-recognition caused by inattentional blindness \cite{johansson2005failure}, mis-understanding, being caught in phishing traps.
Malicious usage includes compliance violation such as unauthorized service usage and access to inappropriate materials.

Such knowledge is provided by the Researcher and Product \& Service Provider, and is then organized and classified by the Registrar.

\subsubsection{Countermeasure Knowledge Base}\label{Sec:CountermeasureKnowledgeBase}

The Countermeasure Knowledge Base is a knowledge base that accumulates information on countermeasures to cybersecurity risks, and itself contains two knowledge bases: Assessment Knowledge Base and Detection/Protection Knowledge Base.

The Assessment Knowledge Base accumulates known knowledge for security level assessment on IT assets.
For instance, rules and criteria for such assessment and the checklists of configurations are accumulated.
Especially, the best practices for such information stored here is quite useful from the viewpoint of reutilzation.
The CVSS/CWSS formula \cite{CVSS,CWSS} is one of the best practices for assessing security levels and is accumulated in this knowledge base.
The eXtensible Configuration Checklist Description Format (XCCDF) \cite{XCCDF} and Open Vulnerability and Assessment Language (OVAL) \cite{OVAL} can be utilized to describe the rules and checklists.

The Detection/Protection Knowledge Base accumulates known knowledge for detecting/protecting security threats.
For instance, rules and criteria for such purpose --- such as intrusion detection system (IDS)/intrusion prevention system (IPS) signatures and detection/protection rules that follow the signatures --- are accumulated.

Such knowledge is provided by the Researcher and Product \& Service Provider, and is then organized and classified by the Registrar.

\subsubsection{Product \& Service Knowledge Base}

The Product \& Service Knowledge Base is a knowledge base that accumulates information on products and services.
It includes two knowledge bases: the Version Knowledge Base and Configuration Knowledge Base.

The Version Knowledge Base accumulates version information on products and services, including naming and enumeration of their versions.
Regarding the product, security patches are also included here.
Common Platform Enumeration (CPE) \cite{CPE} can be utilized in order to enumerate common platforms.

The Configuration Knowledge Base accumulates configuration information on products and services.
Regarding product configuration, it includes naming, taxonomy and enumeration of known configurations.
Common Configuration Enumeration (CCE) \cite{CCE} can be utilized to enumerate common configurations.
Regarding service configuration, it includes guidelines of service usages.

Such knowledge is provided by the Researcher and Product \& Service Provider, and is then organized and classified by the Registrar.

Based on the above discussion, Figure \ref{ProposedOntology.eps} describes the proposed ontology.
It depicts cybersecurity operation domains, entities, cybersecurity operational information and their relationships.

\section{Discussion on Cybersecurity in Cloud Computing}\label{Sec:Discussion}

The proposed ontology is capable of mapping major cybersecurity information standards as shown in Table \ref{Tabl:MajorCybersecurityInformationStandards}.
Based on the proposed ontology shown in Figure \ref{ProposedOntology.eps} and corresponding cybersecurity information standards in Table \ref{Tabl:MajorCybersecurityInformationStandards}, 
necessary information for cybersecurity in cloud computing and the need for standard description formats for such information are discussed for each of the databases and knowledge bases in the following subsections.

\subsection{User Resource Database}\label{Sec:DiscussionUserResourceDatabase}

We define user resources as those that users can utilize regardless of the physical location; be it either in a local IT asset or in the cloud.
Therefore, subscribing cloud services can also be regarded as user resources.
From this viewpoint, the User Resource Database needs to store two additional types of information for cloud security: cloud service subscription information and resource dependency information.
Moreover, security level information such as security level evaluation results should be reviewed to accommodate cloud computing.
The three issues mentioned above are discussed in the following subsections.

\subsubsection{Cloud Service Subscription Information}\label{Sec:CloudServiceSubscriptionInformation}

Major components of Cloud Service Subscription Information are the cloud resource list, data access control policy and identity information.

An Administrator needs to maintain the list of cloud resources to which its organization subscribes, which includes data, applications, hardware and services.
The list is information that may be shared with the other internal organizations.
For instance, an system administrator from a company's headquarters needs to monitor the compliance of its branches.
Or this person needs to deploy the same cloud services in each branch to maintain a uniform IT environment and a uniform security level at each branch.
This information may also be shared with external organizations.
For instance, some products and/or services may be configured automatically based on the subscription list so that they may work effectively and efficiently.
As mentioned above, the list will be shared among different organizations, be it either internal or external, and it is expected to be automatically handled by machines.
Therefore, the description format of the list needs to be standardized so that it can be machine-readable.

The data access control policy defines the rights of user data access, such as those of reading, writing, and executing.
In non-cloud computing, the policy is stored in the local file system.
In cloud computing, it needs to be stored explicitly, and independently from the data and the file system so it can be exchanged.
For instance, eXtensible Access Control Markup Language (XACML) \cite{XACML} can be applied to describe the policy.
In addition to the access control policy for the data itself, that for the cloud location should also be stored and implemented.

Another major component for cloud service subscription information is identity registration.
One can have multiple identities in cyber society.
A list of user identities and identity service registration information needs to be stored.

The other service subscription information such as contract information of each subscribing service should also be stored here.

\subsubsection{Resource Dependency Information}\label{Sec:ResourceDependencies}

Resource dependency information is greatly important in cloud services from the viewpoint of cybersecurity. 

\begin{figure}[htb]
\begin{center}
\includegraphics[scale=.415]{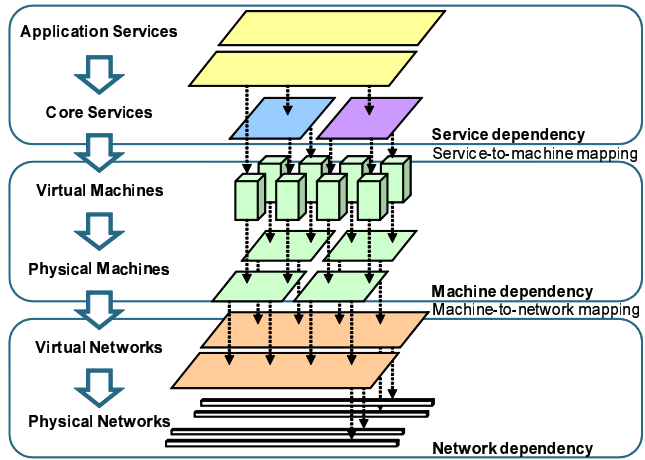}
\caption{Concept of resource hierarchy}
\label{ResourceDependency.eps}
\end{center}
\end{figure}

Different from the non-cloud computing, resources impose very complex dependencies between each other in the cloud computing since they are multi-layered.
In other words, one resource may be built upon another, which can even be built upon another.
The concept of resource hierarchy is described in Figure \ref{ResourceDependency.eps}, which shows resources built upon other resources.
Albeit only six layers are depicted in Figure \ref{ResourceDependency.eps} for ease of understanding, there may exist more hierarchies, which cause more complicated dependencies between resources in a practical environment.
Therefore users may utilize resource A, which utilizes resource B.
In this case, the users are usually not aware of indirect usage of resource B.
Sometimes resource B can be in the cloud.
Under this multi-layered resource environment, damage of a certain resource would affect the other resources directly or indirectly utilizing it.

Therefore the Administrator needs to understand how the damage to one IT resource would affect others.
Clear understanding of resource dependencies facilitates the Administrator to run cybersecurity operations and maintain availability of IT functionality.
Based on resource dependency information between two resources, a dependency graph that describes the chain of resource dependencies can be created.
This graph helps the Administrator immediately recover from failures and drastically improve availability of IT functions.
Provided that the resource dependencies are unclear, and a security flaw is found in one cloud service, the Administrator in a user organization needs to take significant time to identify which IT resources were affected.
The time required to identify the dependency allows the damage to spread further, and causes unavailability of IT services in the user organization.
With the description of resource dependencies, the administrators may run cybersecurity operations efficiently and effectively, and service disruption time is minimized.

Indeed, resource dependency information would provide a great deal of assistance even for non-cloud applications.
Nowadays, virtualization of software is becoming increasingly common, and the emergence of cloud computing expedites the virtualization trend.
In this era, dependency description is of great assistance to cybersecurity and service availability.

Until now, cybersecurity operations have emphasized how to protect systems from external attack.
Yet, from here forward, cybersecurity operations need to also consider service continuity, i.e., service availability.
In order to minimize service disruption at the time of an incident, dependency information is vital.

If Administrators and/or IT Infrastructure Providers have a clear understanding of resource dependencies, it is relatively easy to manage availabilities since most resources are currently managed over virtual systems.
Damaged resources can immediately be replaced with undamaged resources, thus users can enjoy the same IT functions/services without disruption. Once service availability is secured, detailed and sometimes time-agnostic investigation can be conducted for damaged resources.
It also expedites automation of Incident Handling operations.
Albeit damage analysis is rather difficult to automate, the replacement of damaged resources can be done automatically provided the dependency information is clearly understood by the resource holders.


The need for building a standard format for describing resource dependency information shall be considered.
As discussed in section \ref{Sec:WarningDatabase}, it is beneficial to have resource dependency information be machine-readable so machines can automatically process it and identify relevant warning information for specific users.
Beyond that, standardizing information descriptions may cultivate users and operators, and let them realize the importance of managing resource dependency information.
Moreover, having built the standard, the resource dependency information can be shared with external organizations.
For instance, one organization may send a query to a security operator in order to obtain effective countermeasures to cyber attacks with resource dependency information.
Albeit the effective countermeasures may differ for each organization, the security operator may narrow down potentially effective countermeasures based on the resource dependency information and can provide more accurate advice to users.




\subsubsection{Security Level Information}\label{Sec:SecurityLevelInformation}

As with the non-cloud computing environment, security level evaluation is necessary.
Existing information description standards, such as ARF/CRF, can be applied to describe the evaluation results of the security level under a cloud computing environment with proper modification.
If necessary, the evaluation results of local resource and the ones of cloud resource can be separately described on these formats.

Security level scores, such as CVSS/CWSS scores, can be applied to a cloud computing environment.
However, evaluation methodologies, such as the CVSS/CWSS formula, are subject to change.
The methodologies are stored in the Countermeasure Knowledge Base.
Therefore the details of the methodology changes are discussed in section \ref{Sec:CountermeasureKnowledgeBase}.

\subsection{Provider Resource Database}

The Provider Resource Database needs to be expanded to accommodate the needs of cloud security.
In particular, it needs to store two types of information --- subscriber identity and service security level --- which are discussed in the following subsections.


\subsubsection{Subscriber Identity Information}

In order to manage cybersecurity, cloud service providers need to know subscribers' identity information.
At the minimum, they store the list of subscribers and their identities.
Since one user may have multiple identities, the mapping between users and their identities needs to also be stored.
Cloud service providers store not only identity information itself but also its associated information, such as the status of each identity, e.g. valid or invalid, and the reputation of each identity.

\eject

\subsubsection{Service Security Level Information}

Cloud service providers need to provide security information on their services.
Two major ways to show their security levels are by providing a security certificate and a security level evaluation result; both of which are authorized by third parties.

A certificate of cloud service security level may certify whether necessary security control is implemented.
Security control includes implementation of secure technology and security operation processes.
For instance, implementation of trusted log technologies and a security preservation operation cycle can be certified.
Due to the volatile nature of cybersecurity, these certificates need to be reviewed and renewed periodically to maintain their effectiveness.
Since this information will be exchanged among cloud service providers and users, it is more convenient to have this certificate format standardized.

Security levels of cloud services may be evaluated, and the results can be reported.
The format of the evaluation result needs to be standardized so it can be automatically machine-processed.
Since security levels change daily, automatic machine processing is necessary.
Albeit the formula and methodologies of evaluation may differ depending on the evaluator and depending on the purpose, a common set of cloud security evaluation methodologies could be developed, and should be updated periodically; and this common set may be standardized.
For building the standard, metrics should be meticulously designed in order to describe the details of security.
They should be reproducible by a third party.



\subsection{Incident Database}\label{Sec:DiscussionIncidentDatabase}

Users wish to preserve ownership rights for their data.
In other words, they wish to preserve the confidentiality, integrity and availability of their own data even if it is in the cloud.
Thus the data must be confidential even to the cloud service provider where the data is located, must not be manipulated by anybody else without permission from the data owners, and once the data is deleted its restoration must not be possible.

The additional needs incur additional cybersecurity operations.
For instance, monitoring of information property rights violations will be required, and any violation of such rights should trigger warning of a security incident.
Based on the operation change, two types of information need to be stored in the database: data provenance \cite{DataProvenance} and a data placement change log.
The existing scheme to describe incident/event information may also change.
These issues are discussed in the following subsections.

\subsubsection{Data Provenance}

Data provenance is a document history of electronic data and records the details of the processes that produced electronic data back as far as the beginning of time or at least the epoch of provenance awareness.
Therefore, any manipulation of electronic data, not only its content but also its metadata such as the access control policy, needs to be logged.
With data provenance, version management/change tracking of electronic data in the cloud is enabled, and more importantly, violation of data ownership rights can be monitored.
No major standards can be found yet to describe data provenance.
This information should, nevertheless, be exchanged between cloud service providers, users and incident response teams.
Therefore standardizing the format would help improve the quality and efficiency of cybersecurity operations.

\subsubsection{Data Placement Change Log}

Data placement change log is a change log of mapping between the logical and physical locations of data.
One of the biggest changes from the non-cloud computing to the cloud computing is the wide implementation of virtualization.
A data center, for instance, implements resource virtualization technologies and divides its resources so that it can host more customers.
In order to maximize the utilization of its resources, the physical location of the data may change dynamically though its logical location stay the same.
In order to track any incident, the location change, i.e., the change of mapping between the logical location and physical location, needs to be logged.
Note that the data placement change log should be recorded not only for the data in the cloud but also for that in the local IT assets.

This data placement change log can be seen as a part of data provenance.
However, it can also be regarded not as a manipulation of the data but as a manipulation or swapping of infrastructure of the data and thus can be regarded as outside the scope of data provenance.
This is a matter of classification and is outside the scope of this paper.

No major standards can be found yet to describe the data placement change log.
This information should, nevertheless, be exchanged between cloud service providers, users and incident response teams.
Therefore standardizing the format would help improving the quality and efficiency of cybersecurity operations.

\subsubsection{Incident/Event Information}

Cybersecurity operations need to record incident/event information.
Standard formats such as CEE and IODEF have been utilized for this purpose as discussed in section \ref{Sec:IncidentDatabase}.
In order to accommodate cloud computing, proper modifications and/or extensions for these standard formats are required to these standard formats.

Existing standards are designed to describe incident information on assets.
Up until now, they have worked well since assets and data are coupled, and it is very rare for them to be decoupled.
However, in cloud computing, data and IT assets are often decoupled.
Therefore, it is necessary to describe a data incident regardless of the IT asset.
In other words, the target of the incident recording may differ in cloud computing.
Nevertheless, the same information format for describing computer events can be utilized.
Therefore, existing standards such as CEE and IODEF can be adapted to cloud security with proper semantics review.


\subsection{Warning Database}\label{Sec:WarningDatabase}

Warnings need to be issued to users who possess or utilize at-risk resources.
Different from the non-cloud computing environment, a user may utilize a cloud service, which may utilize another cloud service that is facing some security risks.
In other words, a user may indirectly utilize a resource that has severe security risks.
Until now, only the direct users of a specific at-risk resource stayed aware.
Under the cloud computing environment, indirect users of at-risk resources should also maintain caution.

One possible scheme to identify indirect users of at-risk resource is utilizing a resource dependency graph that describes dependency information between resources, as mentioned in section \ref{Sec:DiscussionUserResourceDatabase}.
By looking up the resource dependency graph, users may recognize whether warning information pertains to their own security.

Until now, no major standard format has existed for describing warnings.
However, it may be appropriate to build a certain standard format for the warnings.
In order to automatically judge whether any certain warning information has some affect on user security by accessing the resource dependency graph, the warning information needs to be machine-readable; therefore the format must be standardized.


On the other hand, as with the information in the Incident Database discussed in section \ref{Sec:DiscussionIncidentDatabase}, warnings need to pay attention to data incidents that are independent from IT asset incidents.

\subsection{Cyber Risk Knowledge Base}

Cyber Risk Knowledge Base accumulates knowledge on cyber risks.
Under a cloud computing environment, the impact range of vulnerability need to be described.
Vulnerability caused by mis-configuration will also increase.
These two issues are discussed in the following subsections.

\subsubsection{Impact range of vulnerability}

Different from a non-cloud computing environment, a user may utilize a cloud service that may utilize another cloud service facing some security risks as discussed in section \ref{Sec:WarningDatabase}.
Therefore, vulnerability information needs to be reached by an indirect user of the resource that exposes the vulnerability.
Currently, users only keep aware of resources directly related to their own resources.
In order to judge whether certain vulnerability information affects their security, users may let the vulnerability information access their resource dependency graph as mentioned in section \ref{Sec:DiscussionUserResourceDatabase}.

Albeit it is advantageous to let machines automatically judge the relevance of certain vulnerability information, it is also beneficial to have human-readable information that describes the vulnerability's impact range.
One way of describing this impact range is introducing resource layering, with which each vulnerability specifies its impact range.
Current CVE enumerates vulnerability with no resource-layering concept. For instance, system and application vulnerabilities are enumerated in parallel with no categorization.
If resource-layering concepts such as the six layers shown in Figure \ref{ResourceDependency.eps} are introduced, each vulnerability may identify the layers of the impact range.
This information may increase operators' knowledge.
Therefore, CVE can be extended to describe each vulnerability's impact range.
At the same time, the layering concept needs to be shared by building international standards though it must be meticulously designed.



\subsubsection{Configuration vulnerability}

In addition to the change of impact range, the contents of vulnerability information may change in the cloud computing.
Until now, the main focus of vulnerability was the one found in programming code. However, in the cloud computing, the importance of vulnerability caused by configuration will grow.
Cloud services are based on a combination of multiple components.
Therefore, configuration to let the services work takes on a highly important role.
Consequently, it is expected that a greater number of vulnerabilities caused by configuration will be found in the cloud computing.
Albeit current CWE, which is still a work in progress, has categories for configuration vulnerabilities, these categories are immature and need further development.



\subsection{Countermeasure Knowledge Base}\label{Sec:CountermeasureKnowledgeBase}

The Countermeasure Knowledge Base accumulates assessment rules, which include scoring methodologies and checklists.

Scoring methodologies to evaluate the security level, which are represented by CVSS, need to be developed further, as mentioned in section \ref{Sec:SecurityLevelInformation}.
CVSS provides the scoring scheme for security levels, based on which Administrators can prioritize the urgency of security operations on IT assets.
Albeit the work of CVSS is innovative, it is still in its initial stage of development, and its applicability is limited; it cannot be used for scoring a system that consists of more than a single PC.
Moreover, it is not aware of virtual machines.
Even more, the current formula focuses on the security level of IT asset, not on data decoupled from IT asset.
Future versions of CVSS need to consider these issues.
Albeit there exist some activities that tackle the applicability limitation of the current CVSS, such as Common Assurance Metric (CAM) introduced by the European Network and Information Security Agency (ENISA) \cite{ENISA}, work is still in the initial phases.
Regarding CWSS, albeit it is still under development, the same issue mentioned above needs to be considered in order for it to be applied to cloud security.

A checklist for cloud security, which is best described with OVAL and XCCDF, also needs to be developed further.
Albeit both these works are innovative, as with CVSS they are still in early stages of development and faces the same limited applicability.
Future versions of OVAL and XCCDF need to consider the same issues mentioned above.

\subsection{Product \& Service Knowledge Base}

In the cloud computing, the Product \& Service Knowledge Base needs to contain cloud service enumeration and its taxonomy.
Moreover, configuration information stored in this knowledge base needs to be developed further.
The two issues are discussed in the following subsections.

\subsubsection{Cloud Service Enumeration and Taxonomy}\label{Sec:CloudServiceEnumerationAndTaxonomy}

Albeit some standards exist for expressing product information, such as CPE, no standards exist for expressing cloud services.
In order to enumerate cloud services, standard naming rules of cloud service providers and their services need to be built and commonly shared.
Following the increase of the enumeration of cloud services, taxonomy will need to be built in order to improve usability of knowledge.
This taxonomy needs to be commonly agreed upon, and shared, thus standardization is preferred.
These standards can be either an extension of existing standards or newly built standards.

\subsubsection{Configuration of services}

CCE can still be used to describe the configuration issue of cloud security.
Under a non-cloud computing environment, the configuration of one product was enumerated and stored as knowledge.
Nevertheless, under a cloud computing environment, the configuration of multiple resources needs to be enumerated and stored as knowledge.
Cloud services are based on a combination of multiple resources; therefore configuration of one service consisting of several resources is necessary.
Configuration and interoperability of multiple services are also necessary.

\section{Essential Changes in Cloud\\Computing}\label{Sec:EssentialChanges}

Based on the discussion in section \ref{Sec:Discussion}, this section summarizes the essential changes of cybersecurity information in cloud computing.
Major factors that caused the changes are three: data-asset decoupling, composition of multiple resources and external resource usage.

Data-asset decoupling is a characteristics of cloud computing.
Whereas in the non-cloud computing data and assets were tightly coupled, data and assets can be decoupled and manipulated independently in the cloud computing.
Therefore, in order to preserve data ownership rights for users, data provenance and data placement change logs are required.

Composition of multiple resources is another characteristic of cloud computing, wherein multiple resources comprise one service, and users may indirectly use various resources.
This characteristic is advanced and expedited by resource virtualization technologies.
That requires three types of information: resource dependency information, security assessment methodologies, and configuration information of multiple resources.
Resource dependency information is required to identify who is affected by certain cybersecurity risks and to whom certain cybersecurity information such as warnings and vulnerability needs to be delivered.
Security assessment methodologies in cloud computing is, different from those in non-cloud computing, required to be able to assess security levels of multiple resources as one service.
Configuration information of multiple resources is required to let one service consisting of multiple resources, and let multiple services, work effectively and efficiently.

External resource usage is a further characteristic of cloud computing, wherein external resources are, from a user viewpoint,  utilized as internal resources.
Therefore, enumeration of services and its taxonomy, the list of cloud services a user subscribes to, security level assessment information of cloud services, and identities of users need to be provided.

Meanwhile, we found that the paradigm of security is shifting.
Until now, the major focus of security was on how to protect systems from information security risks.
Nevertheless, how to maintain IT service functionality without service disruptions, i.e., availability, is becoming a primary aspect of security.
Cloud computing enables maintaining of availability since a service consists of multiple resources, each of which can be replaced dynamically at the time of service failure.
In order to maintain availability, aforementioned information such as resource dependency needs to be stored and utilized.

\section{Conclusion}\label{Sec:Conclusion}

This paper proposed an ontological approach toward cybersecurity in cloud computing.
We built an ontology for cybersecurity operational information based on actual cybersecurity operations mainly focused on non-cloud computing.
In order to discuss necessary cybersecurity information in cloud computing, we applied the ontology to cloud computing.

Through the discussion, we identified three major factors that affect cybersecurity information in cloud computing: data-asset decoupling, composition of multiple resources and external resource usage.
Based on the changes in cloud computing, we identified the cybersecurity information necessitated by essential changes such as data provenance and resource dependency information.
Moreover, we found that the paradigm of security is now shifting and availability is becoming one of the most important aspects.

By implementing cybersecurity information identified in this paper, quality cybersecurity operations in cloud computing will be achieved, and cybersecurity in cloud computing will be significantly improved.

\section{Acknowledgments}

We would like to thank Anthony M. Rutkowski, the Rapporteur of ITU-T Q.4/17, from Yaana Technologies, Inette Furey from DHS, Damir Rajnovic from FIRST/Cisco Systems Ltd., Robert A. Martin from the MITRE Corporation, Gregg Schudel from Cisco Systems Ltd. and Craig Schultz from Multimedia Architectures for their many helpful comments, and their insightful perusal of our first draft.
We also want to thank Toshifumi Tokuda from IBM, Hiroshi Takechi from LAC and SeonMeyong Heo from BroadBand Security, Inc. for their extremely useful suggestions.



\bibliographystyle{abbrv}

\end{document}